\documentclass[preprint,
amsmath,
aps, prst
]{revtex4-1}

\usepackage{graphicx}
\usepackage{dcolumn}
\usepackage{bm}
\usepackage{hyperref}

\begin{document}


\title{Dependence of the microwave surface resistance of superconducting niobium on the magnitude of the rf field}
\thanks{This work was supported by the US DOE Office of Nuclear Physics}

\author{A. Romanenko}
\email{aroman@fnal.gov}
\author{A. Grassellino}
\affiliation{Fermi National Accelerator Laboratory, Batavia, IL 60510, USA}%

\date{\today}
             
\begin{abstract}
Utilizing difference in temperature dependencies we decoupled BCS and residual components of the microwave surface resistance of superconducting niobium at all rf fields up to $B_\mathrm{rf}\sim110$~mT. We reveal that the residual resistance decreases with field at $B_\mathrm{rf} \lesssim$~40~mT and strongly increases in chemically treated niobium at $B_\mathrm{rf} > 80$~mT. We find that BCS surface resistance is weakly dependent on field in the clean limit, whereas a strong and peculiar field dependence emerges after $120^\circ$C vacuum baking. 
\end{abstract}

\maketitle

Microwave surface resistance $R_\mathrm{s}$ of superconductors has recently attracted a lot of attention due to its importance for superconducting cavities for particle acceleration~\cite{Padamsee_Review_SUST_2001, Hasan_book2}, microresonators~\cite{Zmuidzinas_ARCMP_2012}, cavity QED~\cite{Raimond_RMP_2001}, and single-photon detectors~\cite{Hadfield_Nature_2009}. For conventional s-wave superconductors, a temperature dependence of $R_\mathrm{s}$ is driven by quasiparticle concentration and follows BCS predictions for $\omega \ll \Delta / \hbar$, $T \ll T_\mathrm{c}$. However, in the limit of $T \rightarrow 0$ experimental data revealed that $R_\mathrm{s}$ saturates at a non-zero value, called residual resistance. A very good approximation for the observed surface resistance is typically provided by~\cite{Turneaure_JAP_1968}:
\begin{equation} 
\label{eq:Rs}
R_\mathrm{s}(T) = A \cdot \frac{\omega^2}{T} \cdot \text{exp}\left(-\frac{\Delta}{k T}\right) + R_\mathrm{res}
\end{equation}
where $\Delta$ is the superconducting gap, factor $A$ depends on the superconductor parameters and an electron mean free path $l$, and, by definition:
\begin{equation} 
R_\mathrm{res} \equiv \lim_{T \rightarrow 0} R_\mathrm{s}(T)
\end{equation}

Unlike temperature dependence well-described by weakly-coupled BCS theory~\cite{Turneaure_JAP_1968, Klein_PRB_1994}, the magnetic field dependence of $R_\mathrm{s}$ is much less understood. For the case of the simultaneously applied static $B_\mathrm{dc}$ and weak rf magnetic fields $B_\mathrm{rf} \ll B_\mathrm{c}$ where $B_\mathrm{c}$ is a thermodynamic critical field, $R_\mathrm{s}$ was investigated experimentally in tin~\cite{Pippard_Tin_IV_PRS_1950, Spiewak_PRL_1958, Spiewak_PhysRev_1959, Richards_PhysRev_1962, Lewis_PhysRev_1964}, aluminum~\cite{Budzinski_PRL_V16_1966, Budzinski_PRL_17_1966, Budzinski_PRB_1973}, and tantalum~\cite{Glosser_PhysRev_1967}. The field dependence $R_\mathrm{s}(B_\mathrm{dc})$was found to be very complex, and anomalies such as a negative slope $dR_\mathrm{s}/dB_\mathrm{dc}$ in some regimes were reported. Several theoretical models were proposed~\cite{Bardeen_PR_1954, Maki_PRL_1965, Koch_PRL_1967, Pincus_PR_1967, Garfunkel_PhysRev_1968} to explain the anomalies. However, for the opposite limit of a zero dc field $B_\mathrm{dc}$ and a strong rf field $B_\mathrm{rf}$, no such detailed studies exist. 

Recent advances in superconducting radio frequency (SRF) cavities for particle acceleration revealed a strong and non-monotonic field dependence of $R_\mathrm{s}$ on $B_\mathrm{rf}$ in the wide range of fields. An origin of such dependence remains a critical outstanding problem. The problem is compounded by the absence of a developed theory of microwave surface resistance at high rf fields. Theoretical calculations of Mattis and Bardeen~\cite{Mattis_Surf_Res_1958} and Abrikosov, Gor'kov and Khalatnikov~\cite{Abrikosov_Surf_Res_JETP_1959} are performed in the limit of weak fields $B_\mathrm{rf} \ll B_\mathrm{c}$ and it is not clear if the results can be extended to high fields. SRF cavities though are intended for the high field operation where condition $B_\mathrm{rf} \ll B_\mathrm{c}$ is not satisfied. For example, for the proposed International Linear Collider an operational gradient is planned to be $E_\mathrm{acc}=35$~MV/m, which corresponds to the peak magnetic field $B_\mathrm{peak}\approx 150$~mT on the cavity surface. Currently, SRF cavities are predominantly made of bulk niobium, and therefore mechanisms contributing to the $R_\mathrm{s}$ of niobium and its field dependence are under intense investigations~\cite{Hasan_book2, Benvenuti_PhysicaC_1999, Grassellino_IPAC_12, Barkov_PRST_2012, Romanenko_SUST_Proximity_2013, Romanenko_HF_PRST_2013}. 

\begin{figure*}[htb]
\includegraphics[width=\linewidth]{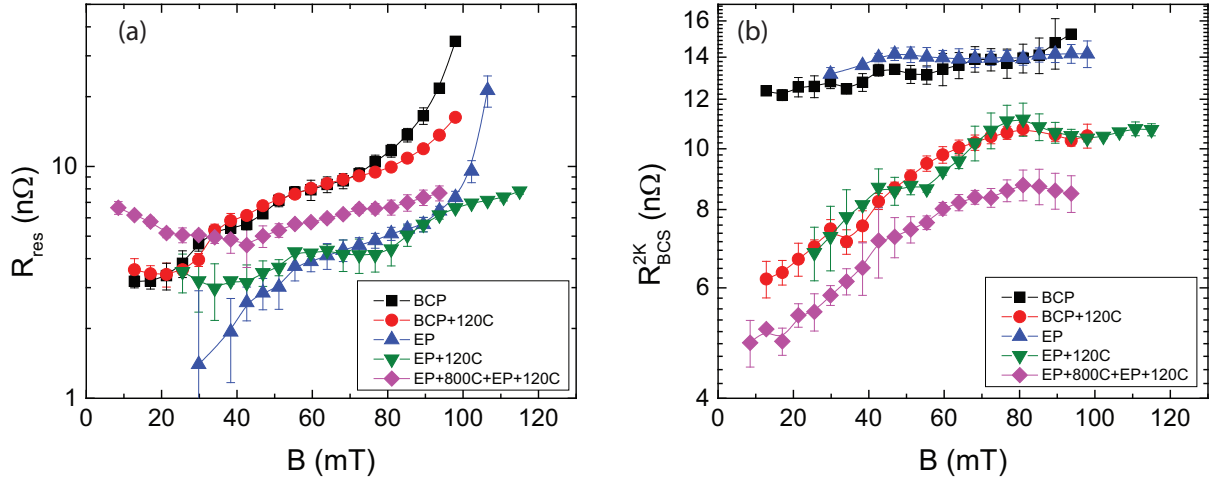}
\caption{\label{fig:R_B} Average (a) residual $R_\mathrm{res}(T \rightarrow 0)$, and (b) BCS surface resistance $R_\mathrm{BCS}(T=2$~K) of niobium as a function of the peak surface magnetic field $B_\mathrm{rf} \equiv B$. BCP = buffered chemical polishing, EP = electropolishing. Lines are to guide an eye and error bars refer to the fit error only.}
\end{figure*}

Three distinct regions of $R_\mathrm{s}(B_\mathrm{rf})$ are typically observed in niobium: so called \emph{low}, \emph{medium} and \emph{high field Q slopes}~\cite{Hasan_book2} discovered in resonant cavities via measurements of the quality factor $Q$, hence the name. The high field $Q$ slope (HFQS) can be removed by mild baking (120$^\circ$C vacuum bake for $\sim$48 hours)~\cite{Kneisel_SRF99} while there is no developed procedures to control low (LFQS) and medium (MFQS) field $Q$ slopes. Physical nature of these $Q$ slopes is not established and is a subject of active research. To study underlying mechanisms and to evaluate proposed models a task of crucial importance is to establish which components of the surface resistance (temperature-dependent $R_\mathrm{BCS}$ or $R_\mathrm{res}$) actually carry which parts of the field dependence. However, up to now, such deconvolution of temperature-dependent and temperature-independent components in $R_\mathrm{s}$ was only done at low fields ($B_\mathrm{rf} < 30$~mT). A single relevant study was reported for niobium films on copper~\cite{Benvenuti_PhysicaC_1999}, which, however, exhibit different microwave behavior from bulk niobium at medium and high fields. 

In this article we report the first explicit deconvolution of the total surface resistance of superconducting niobium into BCS and residual components up to rf surface magnetic fields of $B_\mathrm{rf}\sim110$~mT. Extracted field dependences allow us to significantly advance the understanding of possible underlying mechanisms and, furthermore, provide a long-missing input enabling parametric optimization of SRF cavities at medium and strong accelerating gradients planned for the next generation of particle accelerators. 

For our studies we used bulk niobium (grain size$\sim50 \mu$m, residual resistivity ratio $\sim$300) cavities of TESLA elliptical shape~\cite{TESLA_Cavities_PRST_2000} operating in TM$_{010}$ mode at $f_0 = 1.3$~GHz. For this cavity geometry, surface magnetic field on the majority of the cavity surface is very close to the peak value $B_\mathrm{peak}$, which is proportional to the accelerating gradient $E_\mathrm{acc}$ ($B_\mathrm{peak}/E_\mathrm{acc}=4.26$~mT/MV/m). Cavities were prepared using standard chemical treatments (electropolishing (EP) and buffered chemical polishing (BCP)), and heat treatments such as $120^\circ$C 48 hours and $800^\circ$C 3 hours vacuum bakes. A summary of  surface treatments applied on cavities is presented in Table~\ref{Cavities}.
\begin{table}[htb]
\caption{\label{Cavities}List of cavities used for experiments.}
\begin{ruledtabular}
\begin{tabular}{p{2.5cm}p{6cm}}
Cavity ID & Treatment \\ 
\hline
TE1AES003 & BCP 120 $\mu$m\\ 
TE1AES003 & BCP 120 $\mu$m + 120$^\circ$C 48 hours\\ 
TE1ACC003 & EP 120 $\mu$m\\
TE1ACC005 & EP 120 $\mu$m + 120$^\circ$C 48 hours\\
TE1CAT002 & EP 120 $\mu$m + 800$^\circ$C 3 hours + EP 20 $\mu$m + 120$^\circ$C 48 hours\\
\end{tabular}
\end{ruledtabular}
\end{table}%

Using standard phase-lock techniques~\cite{Knobloch_Cavity_Meas_1991} we measured quality factor $Q(T, B)$ of cavities at several bath temperatures below helium $\lambda$-point ($T_\mathrm{\lambda}=2.17$~K) down to $\sim1.5$~K limited by the cryosystem capabilities. The range of the fields $B$ was limited by either quench or available power. Average surface resistance $R_\mathrm{s}(T, B_\mathrm{rf}) = G/Q(T, B_\mathrm{rf})$ where $G=270$ is a geometry factor determined by the cavity shape, was fitted using Eq.~\ref{eq:Rs} to extract $R_\mathrm{res}$ for each of the cavties. At all fields the quality of fits were excellent with $R^2 \gtrsim 0.99$, and extracted functional dependencies $R_\mathrm{res}(B_\mathrm{rf})$ and $R_\mathrm{BCS}(B_\mathrm{rf})$ at $T=2$~K are shown in Fig.~\ref{fig:R_B}. 

\begin{figure}[htb]
\includegraphics[width=\linewidth]{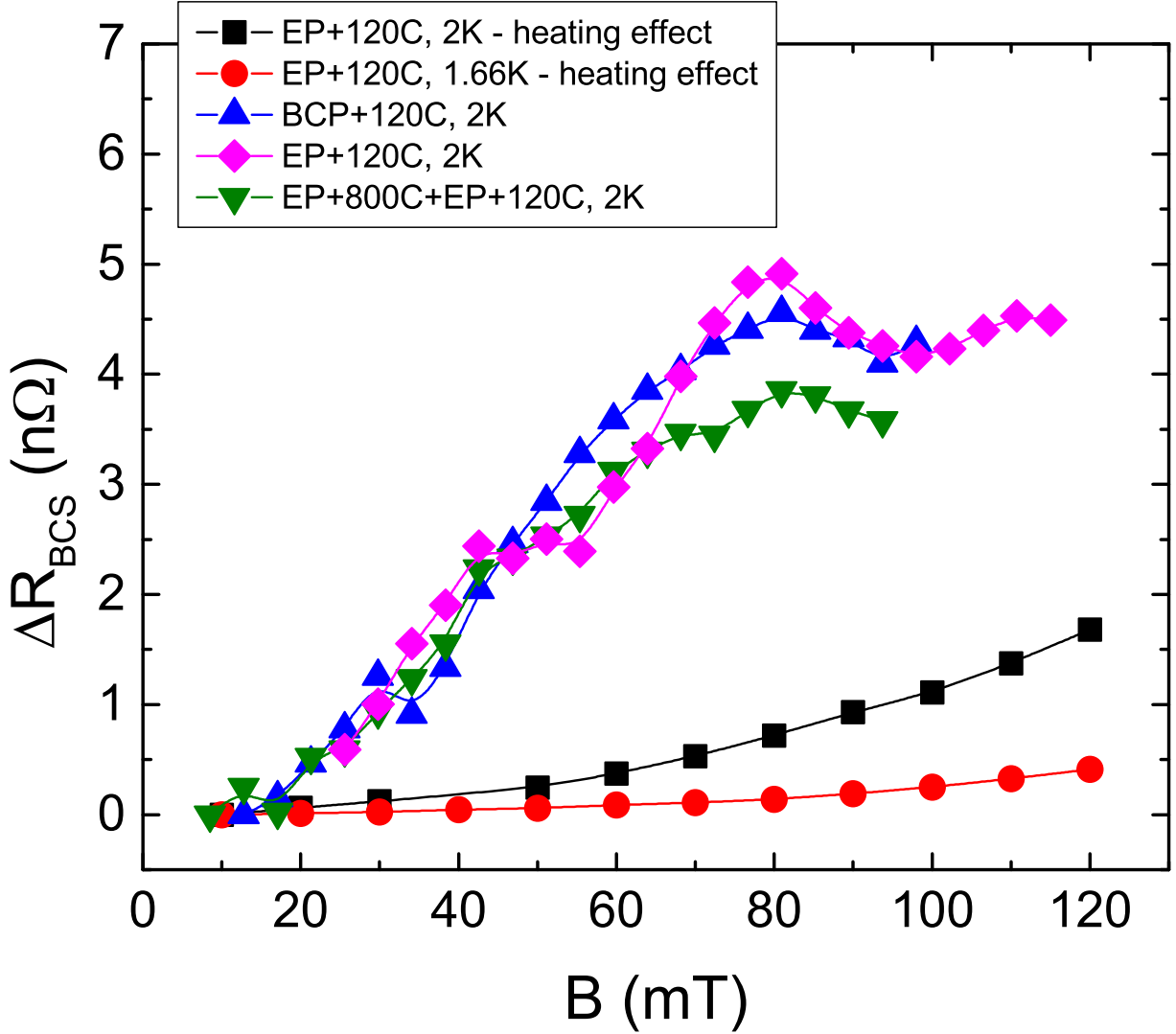}
\caption{\label{fig:Rbcs}Comparison of the observed $R_\mathrm{BCS}$(2~K) increase with field in 120$^\circ$C baked cavities with the heating contribution calculated from thermometry data.}
\end{figure}

In our experiments, the temperature dependence of $R_\mathrm{s}$ was taken versus helium bath temperature $T_\mathrm{b}$. When microwave fields are present in the cavity, the power dissipated in the cavity surface has to be transported by heat diffusion through walls to the helium bath. This causes a non-zero temperature difference $\Delta T_\mathrm{in}=T_\mathrm{rf}-T_\mathrm{out}$ between the inner ($T_\mathrm{rf}$) and outer ($T_\mathrm{out}$) walls of the cavity, and therefore makes $T_\mathrm{rf} > T_\mathrm{b}$. Since $R_\mathrm{BCS}$ is temperature dependent, it is important to consider an effect of this rf surface heating on our results. Fortunately, to estimate its magnitude under different conditions, we can use $\Delta T_\mathrm{out} (B_\mathrm{rf})=T_\mathrm{out}-T_\mathrm{b}$ directly measured on the outside cavity walls by temperature mapping~\cite{Knobloch_T_Map_RSI_1994}. Given the surface magnetic field $H$ and the average surface resistance $R_\mathrm{s}=G/Q_0$ from rf measurements, and the outside wall temperature $T_\mathrm{out}$ from temperature mapping (taking into account thermometers efficiency~\cite{Knobloch_T_Map_RSI_1994} of $\sim35$\%), we apply a 1D heat diffusion model with a surface heat source of $P = 1/2 R_\mathrm{s} H^2$, and niobium thermal conductivity \mbox{$\kappa(T)=0.7\mathrm{e}^{1.65T-0.1T^2}$W/m$\cdot$K}~\cite{Bauer_PhysicaC_2006} to calculate the rf surface temperature $T_\mathrm{rf}$. An expected increase in $R_\mathrm{BCS}$ due to heating is then calculated and the results at $T=2$~K and $T=1.66$~K are shown in Fig.~\ref{fig:Rbcs} in comparison to the experimental $R_\mathrm{BCS}$ increase we found for $120^\circ$C baked cavities. We can conclude that: 1) rf heating is negligible at lower temperatures and has no effect on the extracted $R_\mathrm{res}$; 2) the observed field dependence of $R_\mathrm{BCS}$ cannot be explained by the rf heating. It is important to re-emphasize that rf heating does not introduce any new mechanisms of losses and has no effect on the residual resistance, the only thing it does is make the temperature of the rf surface different from the bath temperature by an amount $\Delta T$, which depends on $B$ in a non-linear way. As we have shown the effect of such $\Delta T$ on our results is negligible.

Our data clearly indicates (Fig.~\ref{fig:R_B}) that both residual and BCS surface resistances carry a field dependence. Several general features are apparent from the data, which we discuss below.

Electropolished niobium without baking exhibits a mild field dependence of the residual resistance up to $B \sim 100$~mT, above which $R_\mathrm{res}$ increases sharply. Baking at $120^\circ$C leads to the elimination of this sharp rise but a slightly increased "residual" slope $dR_\mathrm{res}/dB$ at $B > 80$~mT can still be observed (Fig.~\ref{fig:R_B}(a)). If $800^\circ$C vacuum baking and $20$~$\mu$m removal by electropolishing precedes the 120$^\circ$C bake, an increased slope at $B > 80$~mT is removed as well. 

Buffered chemical polished niobium exhibits a strong dependence $R_\mathrm{res}(B)$.  Similarly to electropolishing, a sharp increase in $R_\mathrm{res}$ at high fields is observed, but with a somewhat lower onset field of $B \sim 80$~mT. Unlike electropolished cavities, baking at $120^\circ$C shifts the onset to slightly higher fields of $B \sim 100$~mT but does not eliminate it. 

Another pronounced feature of the residual resistance is its peculiar field dependence at lower fields $B < 40$~mT with a negative slope $dR_\mathrm{res}/dB < 0$ observed after $120^\circ$C. No such low field features exist in the BCS component.

Baking at 120$^\circ$C has a dramatic effect on the BCS surface resistance and its field dependence. The absolute value of $R_\mathrm{BCS}$ at low fields is decreased by a factor of $\sim2$ in agreement with earlier reports~\cite{Kneisel_SRF99}, and the field dependence switches from very weak to the stronger, remarkably similar among all cavities, field dependence. A slightly lower $R_\mathrm{BCS}$ but with the same field dependence is observed if 800$^\circ$C/light EP treatment is added. As we have shown above, such a field dependence is not due to the rf surface temperature increase and must emerge from the change of the properties of superconducting condensate caused by the magnetic field. Since the decrease in absolute value of $R_\mathrm{BCS}$ is thought to be due to the decrease of the electron mean free path in the magnetic penetration depth~\cite{Kneisel_SRF99} , the change in the field dependence of $R_\mathrm{BCS}$ may also be its consequence. 

Our findings have strong implications for theories attempting to explain the field dependence of $R_\mathrm{s}$ of niobium in different field ranges, and, furthermore, we can provide simple explanations for multiple experimental facts observed in the behavior of niobium $R_\mathrm{s}$.

A mild degradation of $R_\mathrm{s}$ in the medium field range of $30<B_\mathrm{rf}<80$~mT (medium field $Q$ slope) emerges from both $R_\mathrm{res}$ and $R_\mathrm{BCS}$ contributions. Owing to this duality the medium field $Q$ slope is temperature-dependent as well. As the temperature is lowered, although the magnitude of the residual resistance is unchanged, its relative contribution to the $R_\mathrm{s}$ compared to the decreasing with temperature BCS component is increased. Therefore, in those cases where $R_\mathrm{res}$ is stronger field-dependent (BCP, BCP+120$^\circ$C, EP) the slope is enhanced at lower temperatures. For the other treatments (e.g. EP+120$^\circ$C) where the BCS component is stronger dependent on $B$, the situation is the opposite, and lowering the temperature leads to a decrease in the slope. This is exactly what was observed~\cite{Romanenko_PAC_2011} and now can be explained.

A strong increase of the surface resistance at high fields (high field $Q$ slope) is clearly due to the residual resistance increase above a certain threshold (Fig.~\ref{fig:R_B}) and is not an intrinsic property of the superconducting condensate. This finding favors proposed extrinsic mechanisms~\cite{Knobloch_MFE, Romanenko_SUST_Proximity_2013} over intrinsic properties of the quasiparticle spectrum~\cite{Pei_Gurevich_PRB_2012}.

An anomalous low field decrease of $R_\mathrm{s}$ is governed by the corresponding residual resistance changes in contrast to the $NbO_\mathrm{x}$ cluster model~\cite{Halbritter_SRF_2001}, and consistent with the very low breakdown field inclusions model~\cite{Weingarten_PRST_2011}.

An established difference between BCP and EP is different roughness~\cite{Saito_Smoothness_SRF_1993} due to grain boundary steps, while another, less obvious one, is a different vacancy-type defect content~\cite{Romanenko_TFSRF_2012}. Recent $\mu$SR studies~\cite{Grassellino_muSR_PRST_2013} revealed the magnetic flux penetration into BCP cavity cutouts starting at medium fields followed by a sharp increase at higher fields. One of the possible scenarios is that such flux penetration, arguably at sharp grain boundary steps, may explain the difference in $R_\mathrm{res}(B)$ at medium fields between BCP and EP. It was also deduced from the critical exponent analysis~\cite{Casalbuoni_NIM_2005} that surface topology of BCP samples is different from EP ones, and simpler 2D granular current flowing patterns in EP samples are replaced by the more complicated ones in BCP. Such a difference may also play a role in the observed $R_\mathrm{res}(B)$ increase. In order to clearly isolate effects of surface morphology, future experiments are planned with the large or single grain material where only a few grain boundaries are present in the whole cavity.

One of the major findings of the paper is a strong field dependence $R_\mathrm{BCS}(B)$ observed after 120$^\circ$ baking. Previous experiments~\cite{Kneisel_SRF99, Grassellino_TFSRF_2012} suggest that niobium is in the clean limit before and in the dirty limit ($l \ll \xi$) after 120$^\circ$C bake. Therefore, our data suggests that low field $R_\mathrm{BCS}$ calculations~\cite{Mattis_Surf_Res_1958, Abrikosov_Surf_Res_JETP_1959} can be extended to higher fields if the material is in the clean limit. In the dirty limit we observe a significant field dependence $R_\mathrm{BCS}(B)$ with $dR_\mathrm{BCS}/dB > 0$ . A non-linear Meissner effect~\cite{Yip_Sauls_PRL_1992, Groll_NLME_PRB_2010} may produce a quadratic $R_\mathrm{BCS}(B)-R_\mathrm{BCS}(0) \propto B^2$ field dependence. However, our $R_\mathrm{BCS}(B)$ cannot be satisfactorily described by such a law, and the character of the dependence seems to be less steep than quadratic, for which we currently have no satisfactory explanation. 

In conclusion, we experimentally measured for the first time rf magnetic field dependencies at low temperatures of BCS and residual microwave surface resistance in superconducting niobium in the entire range up to the strong rf fields of $B_\mathrm{rf}\simeq$110~mT. We demonstrate that both BCS and residual surface resistances are field-dependent, which allows to explain seemingly complex effects such as a temperature variation of low, medium, and high field $Q$ slopes in SRF cavities for particle acceleration, and enables parametric optimization of future accelerators based on the measured temperature and field dependencies rather than approximate model extrapolations of low field values. Furthermore, our data puts strict constraints on theoretical models proposed to explain the field dependence $R_\mathrm{s}(B)$ in different field ranges.

The authors would like to acknowledge the help and useful discussions of A. Crawford, D. Sergatskov, D. Bice, A. Rowe, M. Wong-Squires, J. P. Ozelis, F. Barkov, O. Melnytchouk and A. Sukhanov. Fermilab is operated by Fermi Research Alliance, LLC under Contract No. De-AC02-07CH11359 with the United States Department of Energy.

\end{document}